\documentstyle[aps,preprint]{revtex}
\begin{document}
\draft
\begin{title}
{\Large\bf Composite absorbing potentials}
\end{title}
\author{\large J. P. Palao, J. G. Muga and R. Sala}
\address{Departamento de F\'\i sica Fundamental y Experimental, 
Universidad de La Laguna, Tenerife, Spain}

\maketitle
 
\begin{abstract}
The multiple scattering interferences due to the
addition of several contiguous potential units
are used to construct composite complex potentials   
that absorb at an arbitrary  
set of incident momenta or for a broad momentum interval.\\  
\end{abstract}


Complex absorbing potentials are an important tool in  
stationary or time dependent scattering calculations
\cite{BKS}. They avoid spurious edge effects of the
finite ``box'' where the 
system is enclosed for numerical purposes in wave packet
calculations \cite{KK}.
They have been also used in other contexts, such as 
time-independent approaches to reactive scattering
\cite{NB90,Neu95,PZ}, or 
calculations of  the microcanonical cumulative
reaction probability, and transition probabilities  in time dependent
fields \cite{SM,PM}.
Allcock, and other authors have claimed that no perfectly absorbing
potential can exist, even for one single incident
momentum, with finite spatial support \cite{All}. But
counterexamples have been found, i.e., potentials that absorb
perfectly at one selected momentum for an arbitrarily small support
do exist \cite{JPA}. In actual
collisions, however, the wave packets may have broad translational
momentum distributions and be peaked around  different momenta.
Such a case arises, for example, as a result of internal state
energy discretization of the collision products 
\cite{Levine}. So the challenge is to model complex
potentials that absorb perfectly at a discrete set of energies, or 
with sufficient efficiency for broad momentum intervals. 
            
In this letter we shall make use of the interferences between paths
associated with ``multiple collisions'' in composite barriers 
to construct complex potentials that absorb at a selected
set of incident momenta or for a broad momentum range. 
Two different complementary methods are described and demonstrated.  
The first method works by successive addition of one potential unit 
for each absorbed momentum and leads to perfect absorption for the 
selected momenta.
In the second method no perfect
absorption is achieved but it is numerically much more robust than
the former, and allows for efficient absorption in a broad momentum
range. 

The first construction method is now described.    
Assume two complex potential units $V_1$ and $V_2$ of contiguous, finite
supports,
$V_1$ being the first one from the left, and $V_2$ the second.  
Let $T_{i}^{r,l}$ and $R_{i}^{r,l}$, $i=1,2$,
be the complex transmission and reflection amplitudes for left ($l$)
and right ($r$) incidence for the $i$-th (isolated)
potential unit, and $T^{r,l}$ or $R^{r,l}$ (without subscript) the 
corresponding amplitudes for the compound barrier, $V=V_1+V_2$.
These last quantities may be obtained in terms of the former 
by the ``multiple collision'' technique \cite{MC}, i.e., by considering
the sum of the amplitudes for all the possible ``paths'' that lead 
eventually to transmission or reflection.
In particular,
\begin{eqnarray}
\label{mc}
T^l&=&T_1^l\left[\sum_{n=0}^\infty(R_2^l R_1^r)^n\right]T_2^l
=T_1^l\frac{1}{1-R_2^lR_1^r}T_2^l
\nonumber\\
R^l&=&R_1^l+T_1^lR_2^l\left[\sum_{n=0}^\infty(R_2^lR_1^r)^n\right]
T_1^r 
=R_1^l+T_1^lR_2^l\frac{1}{1-R_2^lR_1^r}T_1^r\,.         
\end{eqnarray}
Now assume that
\begin{equation}\label{k1}
T_1^l(k_1)=R^l_1(k_1)=0\,,
\end{equation}
where $k_1>0$ is a particular (dimensionless) wave number.
($k=d_1p/\hbar$, where $p$ is the dimensional
momentum and $d_1$ the dimensional width of the first barrier).
Inserting (\ref{k1}) into (\ref{mc}) gives
$T^l(k_1)=R^l(k_1)=0$.
In other words, {\it if $V_1$ is a perfect absorber at $k_1$, the total  
potential $V$ is a perfect absorber for $k_1$ too}.  
The objective of adding $V_2$ is to absorb {\it also} at $k_2\ne k_1$.
If $V_2$ is naively constructed in such a way that
$R_2^l(k_2)=T_2^l(k_2)=0$, according to the equations (\ref{mc})
the composite barrier does not transmit the new momentum,  
$T^l(k_2)=0$, but in general $R^l(k_2)\ne 0$
because of reflection in   
the first barrier, $R^l_1(k_2)\ne 0$.
Instead, if the second potential satisfies 
\begin{eqnarray}\label{T2}
T^l_2(k_2)&=&0\,,\\
\label{R2}
R_2^l(k_2)&=&\frac{R_1^l(k_2)}
{R_1^l(k_2)R_1^r(k_2)-T_1^l(k_2)T_1^r(k_2)}\,,
\end{eqnarray}
total absorption is achieved at $k_2$, $T^l(k_2)=R^l(k_2)=0$.
Eq. (\ref{R2}) is obtained by assuming $R^l(k_2)=0$ and solving for 
$R_2^l$ in (\ref{mc}) in terms 
of quantities that depend only on $V_1$.
Even though the two barriers
alone have non vanishing reflection amplitudes for $k_2$,
when they are put together
the interference effects exactly cancel the global reflection.

We shall next describe a way to construct $V_1$ and $V_2$ with the
desired partial reflection and transmission amplitudes.        
Dimensionless variables will be used throughout. 
The dimensionless position $x$ is obtained by dividing the 
dimensional position by a reference length ($d_1$ in this case),
so that the support of $V_1$ has length one when using the
$x$ variable. The second potential unit, $V_2$, 
may have a different length, $L_2=d_2/d_1$.
Dimensionless energies (kinetic and 
potential) are obtained by dividing the corresponding dimensional 
quantities by $\hbar^2/(2md_1^2)$, $m$ being the mass of the particle.
 
Potentials $V_1$ with vanishing reflection and transmission
coefficients for $k_1$ can be constructed by means of an inversion 
procedure similar to the one described in \cite{JPA}.
The important difference with respect to that work is that now
transmission is allowed at {\it other} wave numbers, so that the wave
function and its derivative are continuous at $x=1$. This is necessary to
take advantage of interference effects.
The boundary conditions to
be satisfied by the stationary wave function corresponding to an 
incident plane wave with wave number $k_1$ at the 
two potential edges are
\begin{eqnarray}\label{four}
\psi_1(0)&=&1,\;\;\;\psi_1'(0)=ik_1,
\\
\nonumber
\psi_1(1)&=&\psi_1'(1)=0\,,
\end{eqnarray}
where the prime means ``derivative with respect to $x$''.
To satisfy these four conditions the wavefunction 
between $0$ and $1$ is written in terms of a functional form with four 
free parameters. By substituting this expression in
the four equations  (\ref{four}), the four parameters are 
determined, and solving in the Schr\"odinger 
equation for $V_1(x)$ one finds
\begin{equation}
V_1=k_1^2+\psi_1''/\psi_1\,.
\end{equation}
The simplest choice for $\psi_1$ is a polynomial,
$\psi_1=\sum_{j=0,3} a_j x^j$.  
From (\ref{four}) the coefficients $a_j$ are readily
obtained,   
\begin{equation}
a_0=1,\;\;\;a_1=ik_1,\;\;\;a_2=-3-2ik_1,\;\;\;a_3=2+ik_1\,.
\end{equation}
To construct the potential $V_2$ between 1 and $1+L_2$, or additional
units with support between two points $x=z$ and $x=z+L$, 
it is convenient to define the new variable $y=(x-z)/L$,
so that the potential unit goes from $y=0$ to $y=1$. A new wavenumber 
is also defined as $\hat{k}=Lk$ (remember that $L$ is dimensionless).
The reflection amplitude $r^l(\hat{k})$ for the new
variables ($\hat{k},\, y$) is related to the one for the original 
variables ($k,\, x$) by $r^l(\hat{k})=R^l(k)e^{-2ikz}$ \cite{RRT}. 
It is much easier to manipulate the constraints (\ref{T2})
and (\ref{R2}) using the new set of variables.
In particular, for obtaining $V_2$, it is first assumed that the 
wave function $\Psi_2(y)$ corresponding to a 
plane wave incident from the
left with wavenumber $\hat{k}_2=L_2 k_2$ obeys the four conditions 
\begin{eqnarray}\label{cond}
\Psi_2(y=0)&=&1+r_2^l\\
\Psi_2'(y=0)&=&i\hat{k}_2(1-r_2^l)\nonumber\\
\Psi_2(y=1)&=&\Psi_2'(y=1)=0\,,
\nonumber
\end{eqnarray}
with $r_2^l(\hat{k}_2)=R_2^l(k_2)e^{-2ik_2}$ and $R_2^l(k_2)$
given by (\ref{R2}).  
By assuming, as before, a polynomial form,
$\Psi_2(y)=\sum_{j=0,3} b_j y^j$,
and substituting in (\ref{cond}), the coefficients
are found to be 
\begin{eqnarray}
b_0&=&1+r_2^l,\;\;\;\;\;\;\;\;\;\;\;\;\;\;\;\;\;\;\;\;\;\;\;\;\;\;\;\;
\;\;\;\;\;\;\;\;\;\,
b_1=i\hat{k}_2(1-r_2^l),\\
b_2&=&-(3+2i\hat{k}_2)-r_2^l(3-2i\hat{k}_2),\;\;\;\;\;
b_3=(2+i\hat{k}_2)+r_2^l(2-i\hat{k}_2)\,.
\end{eqnarray}       
Solving in the Schr\"odinger equation for the corresponding potential 
and rewriting  the result
for the original variables, $\psi_2(x)=\Psi_2(y)$, one then finds
for $V_2$ between $x=1$ and $x=1+L_2$,
\begin{equation}
V_2(x)=k_2^2+\frac{\psi_2''}{\psi_2}L_2^2\,.     
\end{equation}
The potential $V_1+V_2$ so constructed is a perfect absorber
at $k_1$ and $k_2$, 
and addition of a third unit $V_3$ will not change this property,
as we have already discussed for the addition of $V_2$.
By treating the $V_1+V_2$ potential
as a new $\widetilde{V_1}$ unit, and the new barrier $V_3$ as
$\widetilde{V_2}$, 
the inversion method can be repeated to build a potential $V_3$ 
that absorbs $k_3$.  
This procedure can be continued to construct ``perfectly absorbing composite 
potentials'' for an arbitrary number of
momenta . (With little changes the method is also applicable when an
infinite barrier is put at the right edge of the last barrier.
The only difference is that the condition $\psi'(x=\sum L_j)=0$
need not be imposed so that a quadratic polynomial, rather than
a cubic one, is enough for the last barrier.) Note that the polynomials
and the 
minimal set of conditions discussed here have the 
advantage of providing simple explicit expressions but other
functional forms for the wave functions may be used,
and further conditions may be imposed.

Figure 1  shows the survival probability
$S(k)\equiv |R^l(k)|^2+|T^l(k)|^2$ versus $k$ for potentials constructed
with two units in this fashion (thick solid and dotted lines).
The effective
absorption width around $k_2$ increases
with decreasing $L_2$, an effect reminiscent of the the broadening of
transmission resonances when the walls of a double
barrier are narrowed. In Figure 2 three units are used to absorb at
three different wavenumbers.  
The prize to pay for the requirement of perfect absorption is that,
at least for the potential units we have studied, numerical
instabilities complicate their practical implementation. Even though
the potential is explicit, the absorption at an arbitrary
momentum has to be computed numerically. The numerically calculated
$S$ at and around $k_2$ 
becomes very sensitive to small errors in the discretization
of the potentials if $T_1(k_2)$ is extremely small \cite{rl} (It appears as 
a denominator in the estimate of the error of $S$
caused by errors in $R_1^{l,r}$ or 
$T_1$). A way to avoid this 
problem in practice is to truncate the potentials so that $S(k_1)$
is not exactly zero (but sufficiently close for practical purposes,
say $S(k_1)=10^{-5}$). This increases $T_1(k_2)$ and makes $S$ in the
proximity of $k_2$ much more stable with respect to slight numerical
errors due to the discrete representation of the potential.
However, the difficulties increase when adding more $k$ points.
A numerically robust alternative is described next. 

The second method makes also use of interferences 
between contiguous units, but in a less explicit way than the former
approach. 
Now the functional form chosen for the potential is a series 
of $N$ equal length complex square barriers
with complex energies 
$\{V_j\}, \,j=1,2,...,N$. 
The real and imaginary values
of $V_j$
are optimized with standard subroutines 
according to a flexible criterion: The sum of the survivals for a set of 
$s$ selected points is minimized \cite{note},   
\begin{equation}\label{f}
f(V_1,...,V_N;k_1,...,k_s)
=\sum_{\alpha=1}^{s}\,S(V_1,...,V_N;k_\alpha)\,.
\end{equation}
Note that $s$ and $N$ are not necessarily equal.  
In a generic application the $s$ points are evenly spaced in a given 
interval in order to absorb arbitrary wave packets   
within the interval. 
The advantage of this 
functional form for the potential is that, for a given set of 
values $\{V_j\}$, the total transmission and reflection coefficients and
their gradients with respect to variations of the barrier parameters can be 
obtained exactly by multiplication of 
$2\times 2$ transfer matrices. These evaluations are very fast, 
so that many more parameters can be optimized, 
two for each barrier, than for other
functional forms \cite{CPL,CPL2}.     

In Figures 1 and 2 the survival probabilities obtained with the two
methods for $s=2$ and $s=3$ can be compared in the low
(dimensionless) wavenumber region,
which is the important one to minimize the (dimensional)
absorbing potential width 
in the applications. The second method provides with
just a few barriers excellent survival curves
below $0.001$, which is sufficient for most practical purposes.
Note the improvement of the survival
curves as the number of barriers increases. This method achieves
larger absorption widths than a previous
systematic approach \cite{JPA,CPL} in all studied cases \cite{CPL2},
with the added advantage of a very simple implementation.
We have also included in both figures the survival curves for one of the 
most frequently used potentials, $V=-i \eta x^2$,
where $\eta$ (real) has been chosen to minimize $S$
at the two or three selected
points in Figures 1 or 2 respectively. A more extensive comparison 
will appear elsewhere \cite{CPL2}.

We acknowledge many discussions with S. Brouard. 
The work has been supported by Gobierno Aut\'onomo de Canarias (Spain)
(Grant No. PB2/95) and Ministerio de Educaci\'on y Cultura
(Spain) (PB 93-0578).


\newpage
\centerline{FIGURE CAPTIONS}
{\bf Figure 1.} Survival $S(k)$ versus $k$.
The thick solid and thick dotted lines
correspond to ``perfectly absorbing composite
potentials'' with $L_2=0.5$, and $L_2=1.6$ respectively
($L_1=1$ in both cases), that absorb at
$k_1=1$ and $k_2=1.2$.
(For numerical stability the theoretical $V_1$ is truncated
for values of the real or imaginary parts larger than $10^3$.)
The dotted and dashed lines correspond to ``square barrier
composite potentials'' with $N=2$ and $N=3$ respectively,  
with the same total length as 
the potential of the thick solid line, and optimized for the same
values of $k$ ($s=2$).   
The solid line with circles corresponds to the potential 
$-i\eta x^2$, where $\eta$ minimizes the sum of
the survivals at the two selected wavenumbers.

{\bf Figure 2.} $S(k)$ versus $k$.   
The thick solid line corresponds to a   
``perfectly absorbing composite 
potential'' constructed with three units to absorb at $k_1=1.94$,
$k_2=4.84$, and $k_3=7.75$,
$L_1=1$, $L_2=.008$, and $L_3=.024$. As in Figure 1 this potential is
truncated to avoid numerical instability.
Also shown is the survival for square barrier composite potentials
optimized for the same points ($s=3$),
and with the same total length:  
$N=1$, solid line ; $N=2$, dotted line; $N=3$, dashed line;
$N=4$, dotted-dashed line.    
The solid line with circles corresponds to the potential 
$-i\eta x^2$, where $\eta$ minimizes the sum of the survivals 
at the three selected wavenumbers.


\begin{references}
\bibitem{rl} For any isolated potential barrier the left and right 
transmission amplitudes are equal when the potential level is equal on 
both sides.
\bibitem{note} For some numerical applications, a barrier with positive
imaginary part may be problematic. In these cases it is necessary to
minimize $f$ with the constraint ${\rm Im}V_j <0$
\end{references}
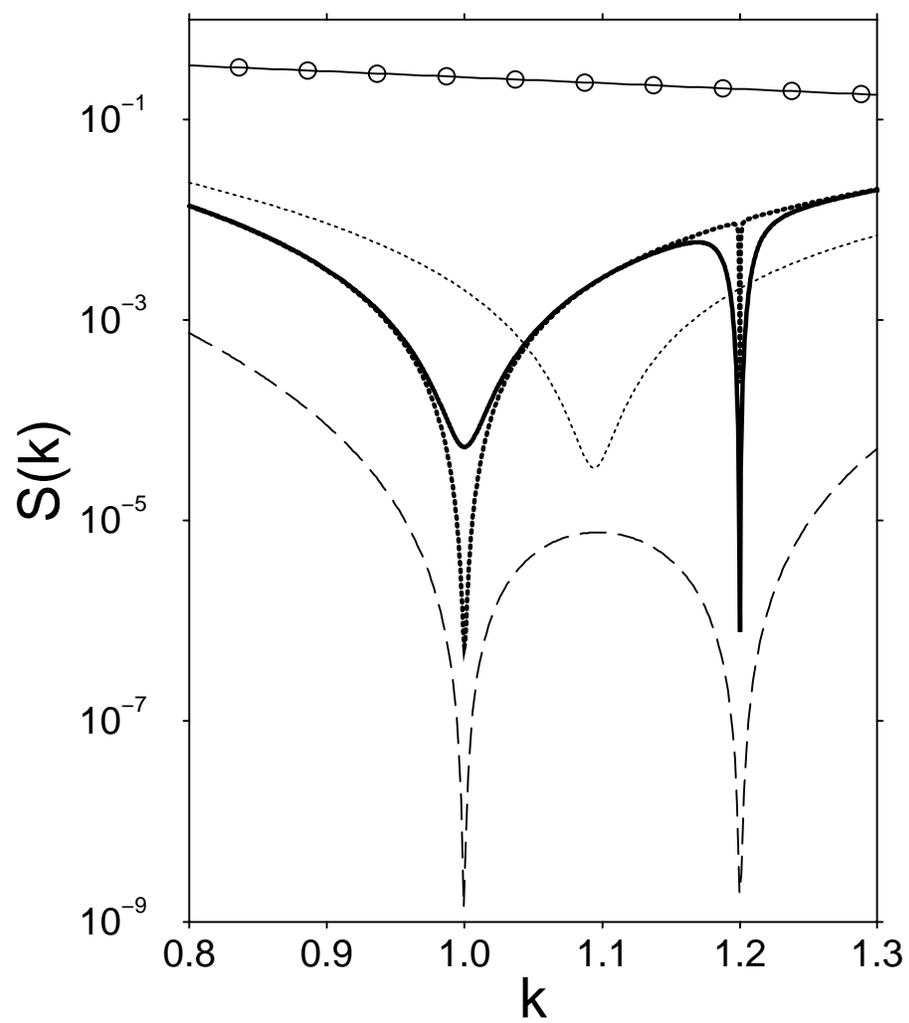
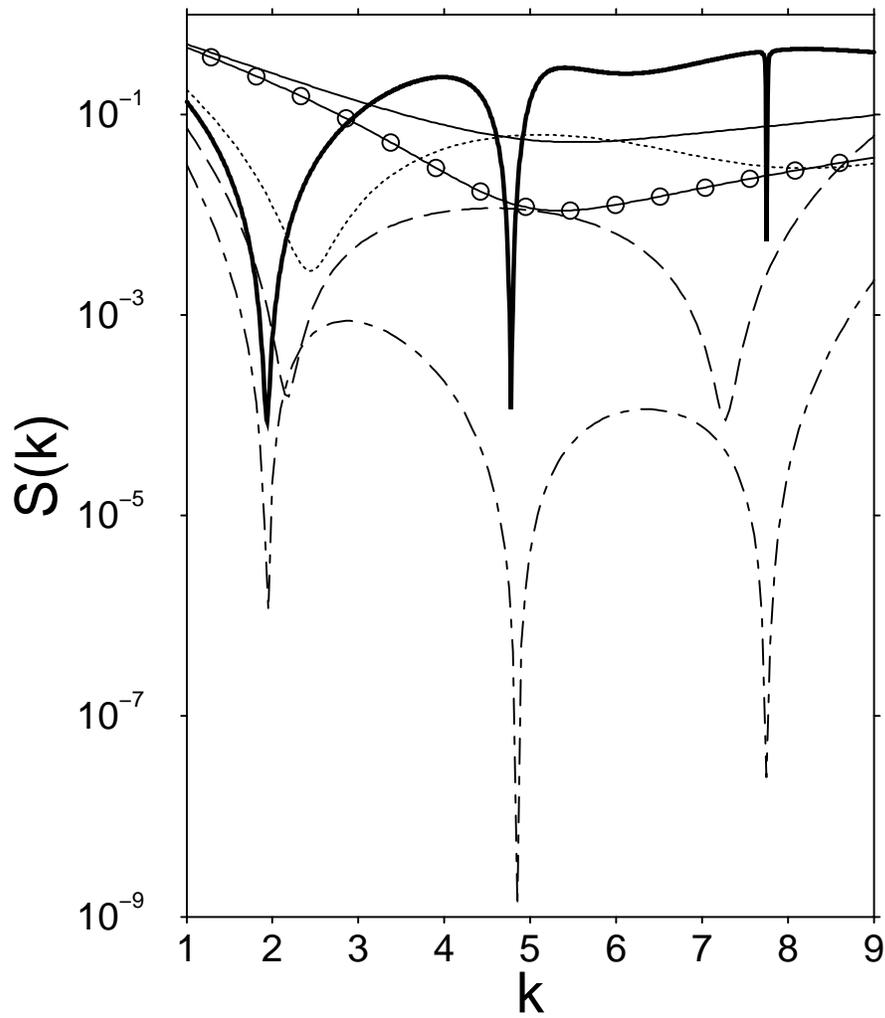
\end{document}